\documentstyle[12pt,epsf]{article}
\catcode`\@=11
\long\def\@makefntext#1{
\protect\noindent \hbox to 3.2pt {\hskip-.9pt
$^{{\ninerm\@thefnmark}}$\hfil}#1\hfill}		%CAN BE USED

\def\@makefnmark{\hbox to 0pt{$^{\@thefnmark}$\hss}}  %ORIGINAL

\def\ps@myheadings{\let\@mkboth\@gobbletwo
\def\@oddhead{\hbox{}
\rightmark\hfil\ninerm\thepage}
\def\@oddfoot{}\def\@evenhead{\ninerm\thepage\hfil
\leftmark\hbox{}}\def\@evenfoot{}
\def\sectionmark##1{}\def\subsectionmark##1{}}
\def\bet{\begin{equation}}
\def\eet{\end{equation}}
\def\bc{\begin{center}}
\def\ec{\end{center}}
\def\ov{\over\displaystyle\strut}
\def\dst{\displaystyle\strut}

\def\Dt{\Delta\tau}

\def\t0{\tau_0}

\def\ben{\begin{eqnarray}}
\def\enn{\end{eqnarray}}
\def\l({\left(}
\def\r){\right)}

\def\t{{\tau}}
\def\o{{out}}
\def\s{{side}}
\def\e{{\eta}}

\renewenvironment{thebibliography}[1]
        {\begin{list}{\arabic{enumi}.}
        {\usecounter{enumi}\setlength{\parsep}{0pt}
	\setlength{\leftmargin 1.25cm}{\rightmargin 0pt}
	\setlength{\itemsep}{0pt} \settowidth
	{\labelwidth}{#1.}\sloppy}}{\end{list}}

\begin{document}

% \leftline{\null  \hfill {\rm hep-ph/9603373}}
\rightline{\null}
\rightline{\null}
\centerline{\normalsize\bf NUMERICAL STUDY OF SPECTRUM AND HBT RADII FOR}
\baselineskip=22pt
\centerline{\normalsize\bf THREE-DIMENSIONALLY EXPANDING}
\baselineskip=22pt
\centerline{\normalsize\bf CYLINDRICALLY SYMMETRIC FINITE SYSTEMS
 }

\centerline{\footnotesize T. CS\"ORG\H O}
\baselineskip=13pt
\centerline{\footnotesize\it Department of Physics, Columbia University,
	538 West 120th, New York, NY 10027}
\centerline{\footnotesize\it MTA KFKI RMKI, H -- 1525 Budapest 114,
POB 49, Hungary}
\baselineskip=12pt
\centerline{\footnotesize E-mail: csorgo@sgiserv.rmki.kfki.hu}
\vspace*{0.3cm}
\centerline{\footnotesize P. L\'EVAI} 
\baselineskip=13pt
\centerline{\footnotesize\it MTA KFKI RMKI, H -- 1525 Budapest 114,
POB 49, Hungary}
\baselineskip=12pt
\centerline{\footnotesize E-mail: plevai@rmki.kfki.hu}
\vspace*{0.3cm}
\centerline{\footnotesize and}
\vspace*{0.3cm}
\centerline{\footnotesize B. L\"ORSTAD}
\baselineskip=13pt
\centerline{\footnotesize\it Physics Institute, University of Lund,
POB 118, S -- 221 00 Lund, Sweden}
\baselineskip=12pt
\centerline{\footnotesize E-mail: bengt@quark.lu.se}
\vspace*{0.3cm}

%\vfill
\vspace*{0.9cm}
\centerline{\footnotesize ABSTRACT}
\vspace*{0.3cm}
{\footnotesize
This study is devoted to a detailed numerical testing of the analytical
results obtained recently for the {\it $M_t$-scaling}
of the parameters of the Bose-Einstein correlation functions.
Numerical testing of analytical results for the spectrum is also performed,
since the geometrical sizes are folded into the momentum distribution
in the scaling limiting case. The analytical results are shown to be
valid in a wider rapidity-range than thought before.
}

%\vspace*{0.6cm}
\normalsize\baselineskip=15pt
\setcounter{footnote}{0}
\renewcommand{\thefootnote}{\alph{footnote}}
\vspace*{1.0 truecm}
\leftline{\bf 1. Introduction}
\vspace*{0.4 truecm}
Recently there has been much interest in the measurement and the 
calculation of the parameters of Bose-Einstein correlation functions (BECF-s)
and those of the invariant momentum distributions 
(IMD-s) for rapidly expanding
systems with flow and temperature profiles.
Such systems are expected to be formed in high energy heavy ion reactions.
The quality and the amount of available data improved drastically
mainly due to the efforts of NA35, NA44, NA49 and WA93 collaborations
at CERN~\cite{na35,na44,na49,wa93}. Data from the dedicated HBT experiment
NA44 indicated an unexpected, scaling behavior for the parameters
of the BECF-s~\cite{na44}. Within the errors of this measurement,
the $R_{side}$, $R_{out}$ and $R_{long}$ parameters of the BECF
for $S + Pb$ 200 AGeV central reactions turned out to be equal and
all were found to scale simultaneously, proportionally to $1/\sqrt{M_t}$, where
$M_t$ is the transverse mass of the identical particle
pair. This scaling relation is
found to be independent of the particle type at the present level of
experimental precision.

This new generation of HBT data triggered a burst of activity mainly from
the Budapest~\cite{nr,1d,halo,qm95,3d,mpd95}, 
Kiev~\cite{sinyu93,hhm:te,akkelin},
Marburg~\cite{marburg,m93} and Regensburg~\cite{uli_sh,uli_l,nix,wiedemann} 
groups, aiming at interpreting the data
in terms of collective, hydrodynamic behavior. The results
from  Los Alamos group ~\cite{losalamos} also indicate collective
behavior caused by re-scattering of secondary particles in the
framework called RQMD event generator. These results are also supported
by simulations of  re-scattering  effects in a hot expanding
gas of hadronic resonances~\cite{humanic}.
 It became clear, that
Bose-Einstein correlations are in general not measuring the whole
geometrical sizes of big and expanding finite systems,
neither in the longitudinal~\cite{sinyukov} nor in the transverse and
temporal directions,
since the expansion may result in strong correlations between space-time
and momentum space variables not only in the longitudinal,
but in the transverse and temporal
directions, too, see ref.\cite{3d} and references therein for a more
detailed account on this topic.

Where have all the geometrical sizes gone?
One can show\cite{1d,3d}, that
they are disguised in the invariant momentum distribution
of the bosons in case they cancel from the
radius parameters of the Bose-Einstein correlation function
(BECF).

We briefly review here the analytical results presented in
refs.\cite{nr,1d,halo,3d,mpd95} which are tested in a detailed
numerical analysis in the subsequent parts. 
The review of the analytical results mainly follows the lines of
ref.~\cite{mpd95}, but the present review is more technical than that,
since we discuss here also the details of the model emission function
and that of the IMD. Ref.~\cite{mpd95} is 
recommended for the illustration of the physical ideas, which determine 
the $M_t$ dependencies of the parameters of BECF-s.

\vspace*{1.0 truecm}
\leftline{\bf 2. Wigner Function Formalism }
\vspace*{0.4 truecm}
The two-particle inclusive correlation function is defined and
approximately expressed in the Wigner function formalism as
\ben
C(\Delta k;K) & = & {\dst \langle n(n - 1) \rangle \ov \langle n \rangle^2}
                {\dst N_2 ({\bf p}_1,{\bf p}_2)
                \ov N_1({\bf p}_1) \, N_1({\bf p}_2) }
                 \simeq
                1 + {\displaystyle\strut \mid \tilde S(\Delta k , K) \mid^2
                                \ov
                        \mid \tilde S(0,K)\mid^2 }.
\enn
In the above line,
the Wigner-function formalism\cite{pratt_csorgo,zajc,chapman_heinz}
 is utilized assuming fully chaotic (thermalized) particle emission.
The covariant Wigner-transform of the source density matrix,
 $S(x,p)$, is a quantum-mechanical  analogue of the classical
 probability that a boson is produced at a given
$ x^{\mu} = (t, {\bf r}) = (t,r_x,r_y,r_z)$
with $p^{\mu} = (E, {\bf p}) = (E, p_x, p_y, p_z)$.
The auxiliary quantity
$ \tilde S(\Delta k , K )  =  \int d^4 x S(x,K) \exp(i \Delta k \cdot x )$
appears in the definition of the BECF,
with $\Delta k  = p_1 - p_2$  and $K  = {(p_1 + p_2) / 2}$.
The single-  and two-particle inclusive momentum distributions (IMD-s)
 are given by
\ben
	N_1({\bf p})  =
	{\dst E \over \sigma_{tot}} {\dst d \sigma \ov d{\bf p}}
		= \tilde S(\Delta k = 0, p),
	\quad & {\mbox{\rm and}} & \quad
	N_2({\bf p_1},{\bf p_2})  =  {\dst E_1 E_2 \over \sigma_{tot}}
                 {\dst d \sigma \ov d{\bf p}_1 \, d{\bf p}_2},
	\label{e:simd}
\enn
where $\sigma_{tot}$ is the total inelastic cross-section.
Note that in this work we utilize the following
normalization of the emission function\cite{halo}:
$\int {\dst d^3{\bf p}\ov E}  d^4x S(x,p) = \langle n \rangle $.

\vspace*{1.0 truecm}
\leftline{\bf 3. Effects from Large Halo of Long-Lived Resonances}
\vspace*{0.4 truecm}
If the bosons originate from a core which is surrounded by
a halo of long-lived resonances, the IMD and the BECF can be
calculated in a straightforward manner. The detailed
description is given in ref.\cite{halo},
here we review only the basic idea. 

If the emission function can be approximately divided into
two parts, representing the core and the halo,
        $ S(x;K)  =  S_{c}(x;K) + S_{h}(x;K) $
and if the halo is characterized
by large length-scales so that     $\tilde S_h(Q_{min};K) $
$ << \tilde S_c(Q_{min};K)$
at a finite experimental resolution of $Q_{min} \ge 10 $ MeV,
then the IMD and the BECF reads as
\ben
        N_1({\bf p}) & = &  N_{1,c}({\bf p}) +  N_{1,h}({\bf p}), \\
        C(\Delta k;K) & = & 1 + \lambda_*
              {\dst \mid \tilde S_{c}(\Delta k, K) \mid^2 \ov
                        \mid \tilde S_{c}(0,K) \mid^2 },
			\label{e:chalo}
\enn
where $N_{1,i}({\bf p})$ stands for the IMD
of the halo or core for $i = h,c$ and
\ben
        \lambda_* & = & \lambda_*(K = p) = \left[{\dst N_{1,c}({\bf p}) \ov
                                 N_{1}({\bf p})} \right]^2.
\enn
Thus within the core/halo picture the phenomenological $\lambda_*$
parameter can be obtained in a natural manner at a given finite
resolution of the relative momentum.
This parameter has been
introduced to the literature by Deutschmann long time ago\cite{deutschmann}.
See ref.~\cite{halo} and references therein for a more detailed account
on the origin of this parameter $\lambda_*$.
In the core/halo picture, the effective
or measured intercept parameter $\lambda({\bf p})$ {\it can be interpreted}
as the {\it momentum dependent} square of the ratio of the IMD
of the core to the IMD of all particles emitted.

\vspace*{1.0 truecm}
\leftline{\bf 4. General Considerations and Results}
\vspace*{0.4 truecm}
We are considering jets in elementary particle
reactions or high energy heavy ion reactions,
which correspond to systems undergoing an approximately
boost-invariant longitudinal expansion.
For expansions fully invariant for longitudinal boosts,
the emission function may depend only on
such variables, which are invariant for longitudinal boosts. These are
defined as $\tau= \sqrt{t^2 - r_z^2}, \, $
 $\eta = 0.5 \ln[(t+z)\,/\,( t-z)],\, $
$m_t = \sqrt{E^2 - p_z^2},\, $
$y = 0.5 \ln[(E + p_z )\,/\,(E - p_z)]\,$
and $r_t = \sqrt{r_x^2 + r_y^2}$.
For finite systems,
 the emission function may depend on $\eta - y_0$
too, where $y_0$ stands for the mid-rapidity.
Approximate boost-invariance is recovered in the
$\mid \eta - y_0 \mid << \Delta y$ region, where
the width of the rapidity distribution is denoted by $\Delta y$.
In terms of these variables the emission function can be rewritten
as
\ben
        S_c(x;K) \, d^4x & = & S_{c,*}(\tau,\eta,r_x,r_y) \,
                                d\tau \, \tau_0 d\eta \, dr_x \, dr_y.
\enn
The subscript $_{*}$ indicates that the functional form
of the source function is changed, and it stands for a
dependence on $K$ and $y_0$ also.

In the standard HBT coordinate system\cite{bertsch},
the mean and the relative momenta are
$K = (K_0,K_{\o},0,K_L)$ and $ \Delta k = (Q_0,Q_{\o},Q_{side},Q_L)$.
Note that the $_{side}$ component of the mean momentum
vanishes by definition\cite{bertsch,3d}.
Since the particles are on mass-shell, we have
$0  =  K \cdot \Delta k = K_0 Q_0 - K_L Q_L - K_{\o} Q_{\o}.$
\begin{figure}
\vspace*{13pt}
          \begin{center}
          \leavevmode\epsfysize=3.0in
       %   \epsfbox{fig1.eps}
          \end{center}
\caption{Emission of particles with a given momentum
is centered around $\tau_s$ and $\eta_s$
for systems undergoing boost-invariant longitudinal expansion,
as indicated by the shaded area.  }
\label{f:1}
\end{figure}
Introducing $\beta_L  = K_L / K_0$ and $\beta_\o = K_\o / K_0$,
the energy difference $Q_0$ can thus be expressed as
\ben
        Q_0 & = & \beta_L Q_L + \beta_{\o} Q_\o. \label{e:11}
\enn
If the emission function has  such a structure that
it is concentrated in a narrow region around
$(\tau_s,\eta_s)$ in the $(\tau,\eta)$ plane,
then one can evaluate the BECF in terms of variables
$\tau$ and $\eta$ by utilizing the expansion
\ben
        \Delta k \cdot x & = & Q_0 t - Q_\o r_x - Q_\s r_y - Q_L r_z \simeq \\
		 \null & \null &
                Q_\t \t - Q_\o r_x - Q_\s r_y - Q_\e \t_s (\eta - \eta_s).
\enn
The coefficients of the $\t$ and the $\t_s (\e - \e_s)$ are new variables
given by
\ben
        Q_\t & = & Q_0 \cosh[\eta_s] - Q_L \sinh[\eta_s] =
	 % \\ \null & \null &
        (\beta_t Q_\o + \beta_L Q_L)  \cosh[\eta_s] - Q_L \sinh[\eta_s], \\
        Q_\e & = &  Q_L \cosh[\eta_s] - Q_0 \sinh[\eta_s] =
	 % \\ \null & \null &
        Q_L \cosh[\eta_s] - (\beta_t Q_\o + \beta_L Q_L)  \sinh[\eta_s].
\enn
In terms of these new variables the BECF reads as
\ben
        C(\Delta k;K) \simeq
                1 + {\displaystyle\strut \mid \tilde S(\Delta k , K) \mid^2
                                \ov
                        \mid \tilde S(0,K)\mid^2 }
                \simeq 		1 + \lambda_*(K)
                {\dst \mid \tilde S_{c,*}
				(Q_\t, Q_\e, Q_\o, Q_\s) \mid^2 \ov
                        \mid \tilde S_{c,*}(0,0,0,0) \mid^2}.
		\label{e:cgen}
\enn
At this level, the shape of the BECF can be rather complicated,
it may have non-Gaussian, non-factorizable structure.
Gaussian approximation to eq.~(\ref{e:cgen}) may break down
as discussed in more detail in the  Appendix of ref.\cite{3d}.

\vspace*{1.0 truecm}
\leftline{\bf 5. Mixing Angle for HBT}
\vspace*{0.4 truecm}
The BECF-s are  frequently\cite{bengt}
but not exclusively\cite{ua1} parameterized by some version of the
 Gaussian approximation.
The out-longitudinal cross-term of BECF has also been discovered in this
context recently\cite{uli_sh}.
In order to identify how this term may come about, let us assume that
\ben
        S_{c,*}(\tau,\eta,r_x,r_y) & = & H_*(\t) \, G_*(\e) \, I_*(r_x,r_y).
        \label{e:fact}
\enn
In Gaussian approximation one also assumes that
\ben
H_*(\tau ) & \propto & \exp(-(\t - \t_s)^2 / (2 \Delta \t_*^2) \, ),
        \label{e:hst} \\
G_*(\eta ) & \propto & \exp(-(\eta - \eta_s)^2 / (2 \Delta\e_*^2) \, ),
        \label{e:gst} \\
I_*(r_x,r_y) & \propto & \exp(-(\, (r_x - r_{x,s})^2 + (r_y - r_{y,s})^2 )
                        / (2 R_*^2) \, ).
\label{e:ist}
\enn
The corresponding BECF is given by a diagonal form as
\ben
C(\Delta k;K) = 1 + \lambda_* \,
                \exp( - Q_\t^2 \Delta\tau_*^2 - Q_\e^2 \t_s^2 \Delta\e_*^2 -
                 Q_t^2 R_*^2). \label{e:c2di}
\enn
This diagonal form shall be transformed to an off-diagonal
one if one introduces the kinematic relations between
the variables $ Q_\t,  Q_\e$ and the variables $Q_\o,Q_L$.
In the HBT coordinate system\cite{bertsch} one finds
\ben
        C(\Delta k;K) & = & 1 + \lambda_* \exp( -R_\s^2 Q_\s^2
			-R_\o^2 Q_\o^2
                  -R_L^2 Q_L^2  -2 R^2_{\o,L} Q_\o Q_L),
	% \\
	% \times \\
        % \null & \null &
        % \times \exp( - 2 R^2_{\o,L} Q_\o Q_L )
	\label{e:crbecf} \\
        R_\s^2 & = & R_*^2, \label{e:rs} \\
        R_\o^2 & = &  R_*^2 + \delta R_\o^2, \label{e:ro} \\
        \delta R_\o^2 & = & \beta_t^2 ( \cosh^2[\e_s] \Delta\t_*^2 +
                                        \sinh^2[\e_s] \t_s^2 \Delta\e_*^2),
                                \label{e:dro}\\
        R_L^2 & = &
        (\beta_L \sinh[\eta_s] - \cosh[\e_s])^2 \t_s^2 \Delta\e_*^2  +
	% \\ \null & \null &
                        ( \beta_L \cosh[\eta_s] - \sinh[\e_s])^2 \Delta\t_*^2,
                        \label{e:rl}\\
        R_{\o,L}^2 & = & (\beta_t \cosh[\eta_s]
        ( \beta_L \cosh[\eta_s] - \sinh[\e_s])) \Delta\t_*^2 + \nonumber\\
                \null & \null &                 (\beta_t \sinh[\eta_s]
                (\beta_L \sinh[\eta_s] - \cosh[\e_s]) ) \t_s^2 \Delta\e_*^2.
                \label{e:rol}
\enn
Note that the effective temporal duration, $\Delta\t_*$
and the effective longitudinal size, $\tau_s\Delta\eta_*$
appear in a mixed form in the BECF parameters $\delta R_\o^2$,
$R_L^2$ and $R_{\o,L}^2$, and their mixing is controlled by
the value of the parameter $\eta_s$.
These results simplify a lot\cite{3d} in the LCMS,
the Longitudinally Co-Moving System\cite{LCMS},
where $\beta_L = 0$:
\ben
        \delta R_\o^2 & = & \beta_t^2 ( \cosh^2[\e_s] \Delta\t_*^2 +
                        \sinh^2[\e_s] \t_s^2 \Delta\e_*^2), \label{e:ldro} \\
        R_L^2 & = & \cosh^2[\e_s] \t_s^2 \Delta\e_*^2 +
                        \sinh^2[\e_s] \Delta\t_*^2, \label{e:lrl} \\
         R_{\o,L}^2 & = & - \beta_t \sinh[\e_s] \cosh[\eta_s] (  \Delta\t_*^2 +
                         \t_s^2 \Delta\e_*^2). \label{e:lrol}
\enn
Let us define the Longitudinal Saddle-Point System (LSPS)
to be the frame where $\eta_s(m_t) = 0$. In LSPS one finds that
\ben
       \delta R_\o^2 & = & \beta_t^2 \Delta\t_*^2, \\
         R_L^2 & = & \t_s^2 \Delta\e_*^2 + \beta_L^2 \Delta\tau^2_*, \\
         R_{\o,L}^2 & = & \beta_t \beta_L \Delta\tau_*^2.
	\label{e:lspsrl}
\enn
Introducing $Q_0 = \beta_t Q_\o + \beta_L Q_L$
and $Q_t = \sqrt{ Q_\o^2 + Q_\s^2}$ the BECF can be rewritten in LSPS as
\ben
C(\Delta k;K) & = & 1 + \lambda_*
                \exp( - \Delta\tau^2_* Q_0^2 - \t_s^2 \Delta\e_*^2 Q_L^2 -
                R_*^2 Q_t^2).
\enn
Thus the out-long cross-term can be diagonalized in the LSPS
frame\cite{nix,3d}. The cross term should be small in LCMS
if $\eta_s^{LCMS} << 1$ i.e. if $\mid y - y_0 \mid << \Delta y$~\cite{3d}.
Since the size of the cross-term is controlled by the value
of $\eta_s$ in any given frame, it follows that
 $\eta_s$ is the {\it cross-term generating
hyperbolic mixing angle}\cite{3d}
for cylindrically symmetric, longitudinally expanding finite
systems which satisfy the
factorization of eq.~(\ref{e:fact}). 
The physical meaning of this hyperbolic
mixing angle $\eta_s$ is illustrated on Figure 1.

\vspace*{1.0 truecm}
\leftline{\bf 6. A New Class of Analytically Solvable Models}
\vspace*{0.4 truecm}
For high energy heavy ion reactions, we model the emission function
of the core with an emission function described in detail
in ref.\cite{3d}. This corresponds to
        a Boltzmann approximation to the local momentum distribution
        of a longitudinally expanding finite system which expands
        into the transverse directions with a transverse flow,
        which is non-relativistic at the maximum of particle emission.
The decrease of the temperature distribution $T(x)$
in the transverse direction is
controlled by  parameter $a$,
the strength of the transverse flow is controlled by  parameter $b$.
         Parameter $d$ controls the strength of the change of the
        local temperature during the course of particle emission\cite{3d}.
If all these parameters vanish, $a = b = d = 0$, one recovers the case
of longitudinally expanding finite systems with $T(x) = T_0$
with no transverse flow, as discussed in ref.\cite{1d},
if $a = d = 0 \ne b$ the model of ref.\cite{uli_sh} is obtained.

Specifically,
 we study the following model emission function for high energy heavy ion
reactions:
\begin{eqnarray}
S(x;K) \, d^4 x & = & {\dst  g \ov (2 \pi)^3}\,  m_t \cosh(\eta - y)\, H(\tau)
		\times \nonumber \\
\null & \null  &
   \exp\l( - {\dst K \cdot u(x) \ov  T(x)} + {\dst \mu(x) \ov  T(x)}\r)
  \,  d\tau \, \tau_0 d\eta \, dr_x  dr_y.
        \label{e:model}
\end{eqnarray}
Here $g$ is the degeneracy factor,
the pre-factor $m_t \cosh(\eta - y)$ corresponds to the flux of the
particles through a $\tau = const$ hyper-surface according to the
Cooper-Frye formula~\cite{cooper} and the four-velocity $u(x)$ is
\ben
u(x) & = &  \l( \cosh(\eta)
        \l( 1+ b^2 \, {\dst r_x^2 + r_y^2 \ov \tau_0^2} \r)^{(1/2)}, \,\,
         b \,{\dst r_x \ov \tau_0}, \,\, b \,{\dst r_y \ov  \tau_0}, \,\,
          \sinh(\eta)
          \l( 1+ b^2 \, {\dst r_x^2 + r_y^2 \ov \tau_0^2} \r)^{(1/2)} \, \r)
          \nonumber \\
\null  & \simeq & \l( \cosh(\eta)
        \l(1 + b^2 \, {\dst r_x^2 + r_y^2 \ov 2 \tau_0^2}\r) ,
        \,\, b \,{\dst r_x \ov \tau_0}, \,\, b \,{\dst r_y \ov  \tau_0}, \,\,
        \sinh(\eta)
        \l( 1 + b^2 \,{\dst r_x^2 + r_y^2 \ov 2 \tau_0^2} \r) \, \r),
        \label{e:4u}
\enn
which describes a scaling longitudinal flow field merged with
a linear transverse flow  profile. The transverse flow is assumed
to be non-relativistic in the region where there is  significant
contribution to particle production.
The local temperature distribution $T(x)$ at the last interaction points
is assumed to have the form
\ben
{\dst 1 \ov T(x)} & = & {\dst 1 \ov  T_0 }
        \l( 1 + a^2 \, {\dst  r_x^2 + r_y^2 \ov 2 \tau_0^2} \r)
           \, \l( 1 + d^2 \, {\dst (\tau - \tau_0)^2 \ov 2 \tau_0^2  } \r),
\enn
and the local rest density distribution is controlled by the
chemical potential $\mu(x)$ for which we have the ansatz
\ben
{\dst \mu(x) \ov T(x) }  & = &  {\dst \mu_0 \ov T_0} -
        { \dst r_x^2 + r_y^2 \ov 2 R_G^2}
        -{ \dst (\eta - y_0)^2 \ov 2 \Delta \eta^2 }.
                \label{e:chmu}
\end{eqnarray}
        The parameters $R_G$ and $\Delta\eta$ control the
        density distribution with finite geometrical sizes.
        The proper-time distribution of the last interaction points
        is assumed to have the following simple form:
\ben
        H(\tau) & = & {\dst 1 \ov ( 2 \pi \Delta \tau^2) ^{(1/2)} }
                        \exp\l( - (\tau - \tau_0)^2 / (2 \Delta \tau^2) \r).
                        \label{e:htau}
\enn
        The parameter $\Delta\tau$ stands for the width of the
        freeze-out hyper-surface distribution, i.e. the emission
        is from a layer of hyper-surfaces which tends to
        an infinitely narrow hyper-surface in the $\Delta\tau \rightarrow 0$
        limit.

 The integrals of the emission function  are evaluated
using the saddle-point method ~\cite{sinyukov,sinyu93,uli_l}.
The saddle-point coincides with the maximum of the emission function,
parameterized by $(\tau_s,\e_s,r_{x,s},r_{y,s})$.
These coordinate values solve simultaneously the equations
\ben
        {\dst \partial S \ov \partial \t } & = &
         {\dst \partial S \ov \partial \e } \, = \,
         {\dst \partial S \ov \partial r_{x} } \, = \,
         {\dst \partial S \ov \partial r_{y} } \, = \, 0.
\enn
These saddle-point equations are solved in the LCMS, the longitudinally
Co-moving system, for $\eta_s^{LCMS} <<1 $ and $r_{x,s} << \tau_0 $.
The approximations are self-consistent if
$ \mid Y - y_0 \mid << 1 + \Delta \eta^2 m_t /T_0 - \Delta \eta^2$ and
 $\beta_t << \tau_s^2 \Delta\eta_T^2 / ( b R_*^2)$ which
can be simplified for the considered model as
 $\beta_t
 = p_t / m_t << (a^2 + b^2) / b/\max(1,a,b) $.
The transverse flow is non-relativistic at the saddle-point if
 $\beta_t  << (a^2 + b^2) / b^2/\max(1,a,b) $.
We assume that $\Delta\tau < \tau_0$ so that the Fourier-integrals involving
$H(\tau)$ in the $0 \le \tau < \infty$ domain can be extended to
the $ -\infty < \t < \infty $ domain.
The radius parameters are evaluated here to the leading order
in $r_{x,s}/\tau_0$. Thus terms of
${\cal O}(r_{x,s}/\tau_0)$ are neglected,
however we keep all the higher-order
correction terms arising from the non-vanishing value of $\eta_s$
in the LCMS. 

 For the model of eq.~(\ref{e:model}) the saddle point approximation for
the integrals leads to an effective
 emission function which
can be factorized similarly to eq.~(\ref{e:fact}).
Thus the radius parameters of the model  are expressible in
terms of the homogeneity lengths
$\Delta\eta_*, R_* $, $\Delta\tau_*$
and the position of the saddle point $\eta_s$
i.e. the cross-term generating
hyperbolic mixing angle.
The saddle-point in LCMS is given by $\tau_s = \t_0$, $\eta_s^{LCMS} =
(y_0 - Y) / (1 + \Delta\eta^2 (1 / \Delta \eta_T^2 -1) )$, $r_{x,s} =
\beta_t b R_*^2 / (\tau_0 \Delta\eta_T^2)$ and  $r_{y,s} = 0$.

\vspace*{1.0 truecm}
\leftline{\bf 7. Geometrical vs. Thermal Length Scales for BECF-s}
\vspace*{0.4 truecm}
        The parameters of the correlation function are related
	by eqs.~(\ref{e:crbecf}-\ref{e:lspsrl}) to
	the parameters $R_*$, $\Delta\tau_*$ and $\tau_s \Delta\eta_*$
	which in turn are given by
\ben
{\dst 1 \ov R_*^2 } & = &
                {\dst 1 \ov R_G^2} +
                {\dst 1 \ov R_T^2 } \cosh[\eta_s],  
		\label{e:rst} \\
{\dst 1 \ov \Delta \eta_*^2} & = &  {\dst 1 \ov \Delta \eta^2 } +
                {\dst 1 \ov \Delta \eta_T^2} \cosh[\eta_s] -
                        {\dst 1 \ov \cosh^2[\eta_s]}, 
		\label{e:dest} \\
{\dst 1 \ov \Delta \tau_*^2 } & = &
                {\dst 1 \ov \Delta\tau^2} + {\dst 1 \ov \Delta\tau_T^2}
                        \cosh[\eta_s].
		\label{e:dtaust}
\enn
Here the geometrical sizes are given by $R_G$, the transverse size,
$\Delta \eta$, the width of space-time rapidity distribution and
$\Delta \tau$, the duration around the mean emission
time $\tau_s = \tau_0$.  The hyperbolic mixing angle $\eta_s\approx 0$
at mid-rapidity $y_0$~\cite{3d}, where also the
out-long cross-term\cite{uli_sh} vanishes.
The thermal length-scales (subscript $_T$) are given by
\ben
 R_T^2  =  { \displaystyle\strut \tau_0^2 \over
        \displaystyle\strut a^2 + b^2 }
         { \displaystyle\strut T_0 \over \displaystyle\strut M_t},
	\quad &
 \Delta\eta_T^2  =  {\dst T_0 \ov M_t},
	& \quad
 \Delta\tau_T^2  =  {\dst \tau_0^2 \ov d^2} {\dst T_0 \ov M_t}.
\enn
The transverse mass of the pair is denoted by $M_t = \sqrt{K_0^2 - K_L^2}$.

These analytic expressions indicate that the BECF views only
a part of the space-time volume of the expanding systems,
which implies that even a complete measurement of the
parameters of the BECF as a function of the mean momentum $K$
may not be sufficient to determine uniquely
 the underlying phase-space distribution.

It is timely to emphasize at this point that the parameters
of the Bose-Einstein correlation function coincide with the
(rapidity and transverse mass dependent)
{\it lengths of homogeneity}\cite{sinyukov}
in the source, which can be identified with that region
in coordinate space where particles with a given momentum are emitted
from. The lengths of homogeneity
for thermal models can be obtained from basically two type of scales
referred to as 'thermal' and 'geometrical' scales.

The thermal scales originate from the factor
$\exp(- p \cdot u(x)  /T(x))$,  where $u(x)$ is the four-velocity field.
This is to be contrasted to the 'geometrical'
scales, which originate from the $\exp( \mu(x) /T(x) )$ factor which
controls the density distribution\cite{3d}. Here $\mu(x)$ stands
for the chemical potential.
How do the changes of the temperature
in the transverse or temporal directions induce transverse mass
dependent thermal radius or thermal duration parameters?
See ref.~\cite{mpd95} for illustration.

As a consequence of possible temporal changes
of the local temperature, we find that the effective duration
of the particle emission $\Delta\tau_*$ may become
transverse mass dependent and for sufficiently large
values of the transverse mass this parameter may become vanishingly small.
The reason for this new effect is rather simple:
Particles with a higher transverse mass are effectively emitted
in a time interval when the local temperature (boosted by the transverse
flow) is higher than the considered value for $m_t$.
If the local temperature changes during the course of particle
emission, the effective emission time for high transverse mass
particles shall be smaller than the effective emission time of
particles with lower transverse mass values.

For a more detailed analysis
of the model the reader is referred to
ref.\cite{3d,japan}, where it is pointed out that under certain
conditions the parameters of the Bose-Einstein correlation
function may obey an $M_t$-scaling: $R_\s \simeq R_\o \simeq R_L
\propto 1/\sqrt{M_t}$.

\vspace*{1.0 truecm}
\leftline{\bf 8. Results for the Invariant Momentum Distributions}
\vspace*{0.4 truecm}
The IMD plays a {\it complementary role}
to the measured Bose-Einstein correlation function ~\cite{1d,nr,qm95}.
Thus a {\it simultaneous analysis} of the Bose-Einstein correlation functions
and the IMD may reveal information
both on  the temperature and flow profiles  and on the geometrical sizes.

For the considered model, eq.~(\ref{e:model}),
the invariant momentum distribution
can be calculated in such a manner, that the Cooper-Frye pre-factor
$m_t \cosh(\eta - y)$ is kept exactly and
the saddle-point approximation is applied to the remaining
Boltzmann and proper-time factors, $\exp( - p\cdot u/T(x) + \mu(x)/T(x) )
H(\tau)$. This calculation yields:
\ben
                N_{1,c}({\bf p}) & = &
        {\dst g \ov (2 \pi)^3 } \,\,
        (2 \pi {\overline{\Delta\eta}_*}^2 \tau_0^2)^{1/2} \, \,
        (2 \pi {\overline R_*}^{\, 2}) \,\, {\dst \overline{\Dt}_* \ov \Dt}\,\,
        m_t  \cosh(\overline{\eta_s}) \, \,
        \exp(+ {\overline{\Delta\eta}_*}^2 / 2) \times  \nonumber \\
        \null & \null &
        \exp( -  p\cdot u(\overline{x}_s )/     T(\overline{x}_s)
                + \mu(\overline{x}_s) /  T(\overline{x}_s) ).\label{e:imdg}
\enn
The  quantities ${\overline{\Delta\eta}_*}^2$ and
$\overline{\eta_s}$
are defined as
\ben
{\dst 1 \ov {\overline{\Delta\eta}_*}^2 }  =
        {\dst 1 \ov \Delta\eta^2} +
        {\dst 1 \ov \Delta\eta_T^2} \cosh[\overline{\eta_s}],
        & \qquad\qquad &
\overline{\eta_s}  = {\dst   (y_0 - y) \ov
        1 + \Delta\eta^2 /  \Delta \eta_T^2 }
\enn
and the modified saddle-point is located in LCMS  at
$\overline{\tau}_s = \tau_s = \tau_0$, $\overline{\eta_s}$,
$ \overline{r_{x,s}} =
\beta_t b \overline{R_*}^2 / (\tau_0 \Delta\eta_T^2)$ and
$\overline{r_{y,s}} = 0$.
The modified radius and life-time parameters can be obtained
by evaluating the $R_*$ and $\Dt_*$ parameters at the
space-time rapidity coordinate of the modified saddle-point,
$\overline{R_*} = {R_*}(\eta_s^{LCMS} \rightarrow \overline{\eta_s})$
and $\overline{\Dt}_* = \Dt_* (\eta_s^{LCMS} \rightarrow \overline{\eta_s})$.
Thus the modified  quantities (indicated by over-line) differ from
the unmodified parameters of the saddle-point approximation
 by the contributions of the Cooper-Frye pre-factor.
 Note, that $\overline{R_*} \simeq {R_*}$ and $\overline{\Dt}_* \simeq {\Dt}_*$
 in the mid-rapidity region, where $\eta_s^{LCMS}, \overline{\eta_s} << 1 $.

In eq.~(\ref{e:imdg}) the exact shape of the four-velocity
field can be used, which is given in the first
line of eq.~(\ref{e:4u}). When evaluating $\mu(\overline{x_s}) /
T(\overline{x_s})$ in any frame, the invariant difference
$\overline{\eta_s} - y_0^{LCMS} =
\overline{\eta_s} + y -  y_0$  should be used.

The momentum distribution as given in eq.~(\ref{e:imdg}) can be rewritten
into a more explicit shape, which is more suitable for analytic study.
This can be done if one neglects
terms of ${\cal O}(\overline{r_{x,s}}^3/\tau_0^3) $ in the exponent,
which is the same order of accuracy which has been utilized for the
solution of the saddle-point equations.
Further, a term in the exponent
$( m_t / T_0) \cosh(\overline{\eta_s}) $
is approximated by its second-order Taylor expansion,
$( m_t / T_0) ( 1 + 0.5 \overline{\eta_s}^2 \, ) $.
These approximations yield
\ben
                N_{1,c}({\bf p}) & = &
        {\dst g \ov (2 \pi)^3 } \,\,
        (2 \pi {\overline{\Delta\eta}_*}^2 \tau_0^2)^{1/2} \, \,
        (2 \pi R_*^2) \,\, {\dst \Dt_* \ov \Dt}
        \,\, m_t  \cosh(\overline{\eta_s}) \, \,
        \exp(+ {\overline{\Delta\eta}_*}^2 / 2) \,\,
        \times  \nonumber \\
\null & \null & 
        \exp\l( \mu_0 / T_0 \r) \, \,
	\exp\l( - {\dst (y - y_0)^2 \ov 2 (\Delta\eta^2 +
        \Delta \eta_T^2) } \r) \,
        \times  \nonumber \\
\null & \null & 
\exp\l( - {\dst m_t \ov T_0} \l(  1 - f \, {\dst \beta_t^2 \ov 2} \r) \r) \,
\exp\l( - f \, {\dst m_t \beta_t^2 \ov 2 (T_0+ T_G)} \r),
\label{e:imd}
\enn
where the geometrical contribution to the effective temperature is given by
$T_G(m_t)  = T_0 \,  R_G^2 / R_T^2(m_t) =
(a^2 + b^2)\, m_t \, R_G^2 / \tau_0^2$ and the fraction $f$ is defined as
$f = b^2 / (a^2 + b^2)$, satisfying $0 \le f \le 1$.
 The following relations hold:
\begin{eqnarray}
\Delta y^2(m_t) & = & \Delta \eta^2 + \Delta \eta_T^2(m_t) \qquad
\mbox{\rm and} \qquad
{\dst 1 \ov T_*}  =  {\dst f \ov T_0 + T_G(m_t=m) } + {\dst 1 -f \ov T_0}.
        \label{e:tempe}
\end{eqnarray}
Thus e.g. the width $\Delta y^2(m_t)$ is dominated by the
longer of the geometrical and thermal length scale, in contrast to 
the HBT radius parameters which are dominated by the shorter of these scales.
That is why the IMD measurements
can be considered to be complementary to the BECF data.

\vspace*{1.0 truecm}
\leftline{\bf 9. Analysis of Limitations} 
\vspace*{0.4 truecm}

The simple analytic formulas presented in the previous sections
are obtained in
a saddle-point approximation for the evaluation of the space-time
integrals. This approximation is known to converge to the
exact result in the limit the integrated function develops a
sufficiently narrow peak, i. e. both
\ben
\Delta\eta_*^2( y, m_t) << 1 
	&\mbox{\rm and} &\Delta \tau_*^2( y, m_t) / \tau_s^2 << 1
	\label{e:mtceta}
\enn 
are required.
These in turn give a lower limit in $m_t$ for the applicability
of the formulas for the class of models presented in the previous section.
These limits were studied analytically in ref. ~\cite{3d},
here we analyze them numerically in the subsequent parts.

Compared to the condition ~(\ref{e:mtceta}), the
condition of validity of the
calculation of the invariant momentum distribution is less
stringent, since one needs to satisfy only $\overline{\Delta\eta}_*^2 << 1$.
In the mid-rapidity region where NA44 data were taken,
one has $\eta_s \simeq 0 $ and for {\it finite} systems
one finds $ m_t >> T_0 (2 - 1 /\Delta\eta^2)$.
Note that this estimated lower limit in $m_t$ is extremely sensitive to
the precise value of $\Delta\eta$ in the region $\Delta\eta \simeq 1 /\sqrt{2}
\approx 0.7$.
For finite systems, the region of applicability of our results
extends to lower values of $m_t$ than for infinite systems which were
recently studied in great detail in ref.~\cite{wiedemann}.
An upper transverse momentum limit
is obtained for the validity of the calculations
from the requirement
\ben
r_{x,s} / \tau_0 & < & 1
		\label{e:xsc}
\enn
 or  $r_{x,s}^2 / \tau_0^2 << 1$.
This condition and the requirement $ b r_{x,s} / \tau_0 < 1$
has to be fulfilled simultaneously.
Finally we note that the linearization of the saddle-point equations
assumes that
\ben
	\eta_s^{LCMS} < 1
\enn
and for the IMD calculations a less stringent condition
$\overline{\eta_s} < 1$ was assumed. 

When comparing to data, detailed numerical studies may be
necessary~\cite{wiedemann} to check the precision of the saddle-point
integration. In the subsequent chapters, we present these tests
for the analytic results given above
and indicate the numerically found  
validity or violation of the conditions given in this Chapter.

\begin{figure}
          \begin{center}
          \leavevmode\epsfysize=6.0in
       %   \epsfbox{f2.eps}
          \end{center}
\caption{
}
BECF parameters in LCMS, the IMD and the essential 'small' parameters.
 Solid line shows the
results of numerical integration, dash-dotted line indicates
the analytical results. Note that the transverse flow is switched off
and $T(x) = T_0$ as follows from $ a = b = d = 0$.
The longitudinal size is chosen to be small, $\Delta\eta = 0.3$.
On the last sub-plot, solid line indicates $\Delta\eta_*$, 
 dash-dotted line stands for $r_{x,s}/\tau_s$ and dashed 
line for $\eta_s^{LCMS}$.
These quantities are required to satisfy $\Delta\eta_*^2 << 1$,
$r_{x,s}^2/\tau_s^2 << 1$ and
$\eta_s^{LCMS} << 1$. The other conditions, $\Delta\tau_*^2 /\tau_s^2 << 
1$ and $\overline{\Delta\eta}^2_* << 1$ are weaker conditions and 
thus not indicated on this and the subsequent plots.
\label{f:2}
\end{figure}

\begin{figure}
          \begin{center}
          \leavevmode\epsfysize=7.0in
     %     \epsfbox{f3.eps}
          \end{center}
\caption{
}
Same as Figure~\ref{f:2}, but 
with increased longitudinal size,  $\Delta\eta = 1.25$.
In the low-$m_t$ region $\Delta\eta_*^2 \ge 1$, which results in a
slight increase of the numerically calculated
 $R_{out}$ component as compared to the analytical result.
\label{f:3}
\end{figure}

\begin{figure}
          \begin{center}
          \leavevmode\epsfysize=6.5in
          % \epsfbox{f4.eps}
          \end{center}
\caption{
}
Transverse flow, transverse and temporal temperature
gradients are switched on. Parameters correspond to a
scaling limiting case at mid-rapidity.
The numerical integration (solid line) is performed for the approximate
transverse flow profile. Dash-dotted line indicates the analytical
results. For the IMD sub-plot, dashed line stands for the simplified
analytical result, eq.~(\ref{e:imd}), and dash-dotted line indicates
the slightly more precise eq.~(\ref{e:imdg}).
\label{f:4}
\end{figure}

\begin{figure}
          \begin{center}
          \leavevmode\epsfysize=7.0in
          %\epsfbox{f5.eps}
          \end{center}
\caption{
}
Same as Figure~\ref{f:4}, but now the numerical integration
was performed with the exact transverse flow profile.
\label{f:5}
\end{figure}

\begin{figure}
          \begin{center}
          \leavevmode\epsfysize=7.0in
          %\epsfbox{f6.eps}
          \end{center}
\caption{
}
Same as Figure~\ref{f:4}, but the longitudinal size
of the system is increased.
\label{f:6}
\end{figure}

\vspace*{1.0 truecm}
\leftline{\bf 10. Numerical test for the a = b = d = 0 case}
\vspace*{0.4 truecm}

Let us start the numerical analysis with a set of parameters,
that satisfy both conditions~(\ref{e:mtceta}) and (\ref{e:xsc}).
This is the reason for choosing $ a = b = d = 0$ and 
$\Delta \eta = 0.3$ which guarantees that $\Delta \eta_* << 1$
is satisfied.
We start the analysis at mid-rapidity, and devote the last chapter
to studies in the target and projectile fragmentation region.
Note that the out-longitudinal cross-term~\cite{uli_sh} vanishes
at mid-rapidity.

Due to the large number of model-parameters a rather
comprehensive analysis is necessary and it is important to 
check the effects from various type of parameters.

As one can see on Figure \ref{f:2} the agreement between the 
analytical results and the numerically evaluated radius parameters
and IMD is almost perfect.
Note that for the analytical calculations of the HBT radii in LCMS
eqs.~(\ref{e:crbecf}-\ref{e:lspsrl},\ref{e:rst}-\ref{e:dtaust}) 
were used, the IMD has been analytically
evaluated both from the slightly more precise
eq.~(\ref{e:imd}) and from the more approximate eq.~(\ref{e:imdg}), 
both yielding results indistinguishable from the numerically integrated 
curve. For the numerical results on the HBT radii,
we have evaluated the model-independent Gaussian radii in LCMS,
as given in refs.~\cite{uli_l,uli_sh}, for the model emission 
function of eq.~(\ref{e:model}).
For the numerical evaluation of the IMD, we have utilized eq.~(\ref{e:simd}).

How significant is the limitation 
given by eq.~(\ref{e:mtceta})? Can we apply the analytical results
to systems created at 200 AGeV  bombarding energy,
where $\Delta \eta \simeq 1.5$? According to Figures \ref{f:3} and
\ref{f:6} the analytical results provide a good approximation
even in this case, (note that Figure \ref{f:6}
is evaluated for $\Delta \eta = 1.5$ for $a, b, d \ne 0$). 
The relative error  on these figures is maximal around
$m_t = 240 $ MeV for pions, being about 10 \% for $\Delta \eta = 1.2$,
20 \% for $\Delta \eta = 1.5$. These errors are characteristic for  
 the $R_{out} $ and 
$R_{long} $ radius parameters. The side radius component and the spectrum
is obtained with much smaller errors. The relative error of the
analytical results vanishes with increasing $m_t$ for this case.
Thus the analytic approximations are reasonably good, sometimes
excellent if $a = b = d = 0$ and $\Delta \eta \le 1.5$.

\vspace*{1.0 truecm}
\leftline{\bf 11. Flow and Temperature Profile Effects }
\vspace*{0.4 truecm}

The cases with non-vanishing parameters $a,b$ and $d$ are studied
in this chapter. Figures \ref{f:4} and \ref{f:5} compare the 
analytical results with the numerical ones for $\Delta\eta = 0.7$
case, Figure \ref{f:6} shows the same for
$\Delta\eta = 1.5$. The model parameters of these figures 
correspond to an
approximate $m_t$ scaling for the radius parameters of the BECF.
On Figure \ref{f:5}, the model was integrated numerically 
utilizing the exact flow profile, as given by the first line
 of eq.~(\ref{e:4u}).
On Figures \ref{f:4} and \ref{f:6}, the approximation 
given in the second line of eq.~(\ref{e:4u}) 
has been performed before the numerical
integration. In this case, the high $m_t$ limit of the analytical
calculations corresponds to that of the numerical one, as it should 
because the radius parameters of saddle-point integration 
decrease with increasing $m_t$ and thus 
the precision of the saddle-point integration increases with
increasing value of $m_t$. 

In the region of $m_t - m \le 200 $ MeV, the relative error of the analytical
calculations for the HBT radius parameters is
about   10 \%. In case of the IMD the relative error
refers to that of the slope parameter, which is approximately 10 \% 
on Figures \ref{f:4} and \ref{f:6}.

It is worth comparing Figure \ref{f:4} with Figure \ref{f:5}.
The analytical curves on these figures are the same, but the
flow profile eq.~(\ref{e:4u}) is exact on Figure \ref{f:5},
while the approximation in eq.~(\ref{e:4u}) was performed before
the numerical integration on Figure \ref{f:4}.
The side and the longitudinal radius parameters are not changed significantly,
however the high $m_t$ part of the spectrum and the out radius parameter
is enhanced, increasing the deviation between numerical and analytical
results to about 20 - 25 \% in the kinematic region $m_t - m < 600$ MeV.
Note that the two analytic approximations for the spectrum,
eqs.~(\ref{e:imdg}) and (\ref{e:imd}) yield slightly different results,
as indicated by the dashed and dash-dotted lines on the IMD part of the
Figures, respectively.

\vspace*{1.0 truecm}
\leftline{\bf 12. Pions vs. Kaons } 
\vspace*{0.4 truecm}

One expects that the precision of the calculation is increased for 
particles heavier than pions, because the thermal scales and thus the
radius parameters decrease with increasing transverse mass.
This is indeed the case, as indicated by Figures \ref{f:7} -- \ref{f:10}.
When evaluating the numerical integrals for these
figures, as well as Figures~\ref{f:11} - \ref{f:16},
 the exact flow profile has been utilized.

The radius parameters are dominated by
the geometrical ones on Figures \ref{f:7} and \ref{f:8},
resulting in a weaker $m_t$ dependence, while 
on Figures \ref{f:9} and \ref{f:10} the HBT radius parameters
are dominated by the thermal scales since $R_G > R_T(m)$ for pions,
and the radius parameters satisfy approximate $m_t$ scaling.
On these figures, the precision of the analytic results
is nowhere worse than 20 \%, in some cases especially for kaons and
for the side and the longitudinal radius parameter at higher $m_t$ 
the relative errors become less than 5 \%.
\begin{figure}
          \begin{center}
          \leavevmode\epsfysize=7.0in
          %\epsfbox{f7.eps}
          \end{center}
\caption{
}
Numerical integration done with exact flow profile,
case $R_G < R_T(m)$ for pions.
\label{f:7}
\end{figure}

\begin{figure}
          \begin{center}
          \leavevmode\epsfysize=7.0in
          %\epsfbox{f8.eps}
          \end{center}
\caption{
}
Same parameters as for Figure~\ref{f:7}, but for kaons.
\label{f:8}
\end{figure}

\begin{figure}
          \begin{center}
          \leavevmode\epsfysize=7.0in
          %\epsfbox{f9.eps}
          \end{center}
\caption{
Numerical integration done with exact flow profile,
case $R_G > R_T(m)$ for pions.
}
\label{f:9}
\end{figure}

\begin{figure}
          \begin{center}
          \leavevmode\epsfysize=7.0in
          %\epsfbox{f10.eps}
          \end{center}
\caption{
}
Same parameters as for Figure~\ref{f:9}, but for kaons.
\label{f:10}
\end{figure}

\begin{figure}
          \begin{center}
          \leavevmode\epsfysize=7.0in
          %\epsfbox{f11.eps}
          \end{center}
\caption{
}
Results for kaons, using a parameter set where 
the limitation ${x}_s(m_t)/\tau_0 << 1$ is violated:
${x}_s(m_t)/\tau_0 \ge 0.6$.
\label{f:11}
\end{figure}

\begin{figure}
          \begin{center}
          \leavevmode\epsfysize=7.0in
          %\epsfbox{f12.eps}
          \end{center}
\caption{
Results for pions, using a parameter set where 
the limitation ${x}_s(m_t)/\tau_0 << 1$ is violated.
}
\label{f:12}
\end{figure}

\begin{figure}
          \begin{center}
          \leavevmode\epsfysize=7.0in
          %\epsfbox{f13.eps}
          \end{center}
\caption{
}
Another violation of the limitation
 ${x}_s(m_t)/\tau_0 << 1$, using $a = 0$ and $b = 1$. 
\label{f:13}
\end{figure}

\begin{figure}
          \begin{center}
          \leavevmode\epsfysize=7.0in
          %\epsfbox{f14.eps}
          \end{center}
\caption{
}
Numerical versus analytical results at $y  = 2$.
Note that mid-rapidity is located at $y_0 = 3$ and that 
the longitudinal size is given by $\Delta\eta = 1.5$.
Note that $\Delta\tau = 0$ fm/c.
\label{f:14}
\end{figure}

\begin{figure}
          \begin{center}
          \leavevmode\epsfysize=7.0in
          %\epsfbox{f15.eps}
          \end{center}
\caption{
}
Same as Figure~\ref{f:14}, but in the target fragmentation region
$y = 1$ and for $\Delta\tau = 2 fm/c$.
\label{f:15}
\end{figure}

\begin{figure}
          \begin{center}
          \leavevmode\epsfysize=7.0in
          %\epsfbox{f16.eps}
          \end{center}
\caption{
}
Same as Figure~\ref{f:15}, for $y = 6$.
\label{f:16}
\end{figure}

\vspace*{1.0 truecm}
\leftline{\bf 13. Beyond the Limitations} 
\vspace*{0.4 truecm}

We have already seen that the conditions for the smallness
of $\Delta\eta_*$ are automatically satisfied at higher $m_t$
thus the requirement (\ref{e:mtceta}) is not very important. 
On the previous graphs, we have shown figures where
the other condition for the validity of the calculation,
eq.~(\ref{e:xsc}) has been satisfied.
In this chapter we violate this condition eq.~(\ref{e:xsc})
by a seemingly innocent change in the values of the parameters
$ a$, $b$, $R_G$ and $\tau_0$. 
This way the saddle-point $x_s(m_t)$ deviates more and more
from zero and it moves outside the region where the
linearization of the saddle-point equations is allowed.
Thus the analytical approximations break down as illustrated on
Figures \ref{f:11} -- \ref{f:13}.
The most serious defect is observable in the sub-plots indicating
the IMD. The slope-parameter is very badly estimated
by the analytical results if the condition ~(\ref{e:xsc}) is violated.
A good numerical interpretation of the condition ~(\ref{e:xsc}) 
can thus be formulated as $r_{x,s} / \tau_0 \le 0.6$.
(Here we assume $b \le 1$.) 
If $r_{x,s} / \tau_0 \le 0.6$, the numerical and analytical results
are in 10 -20 \% agreement at mid-rapidity, however if
$r_{x,s}^2 / \tau_0^2 > 0.5$, the reliability of the analytic calculation
is lost. The precision of model calculations for pions are very
sensitive to whether this condition is well satisfied or violated. 

Note that our analysis has been performed with $T_0 = 140 $ MeV fixed.
Lower values of $T_0$ improve
the agreement between the analytical results and the 
numerical ones, while larger values of $T_0$ in general
make the deviations larger in a given kinematic region.

Surprisingly, the HBT parameters for kaons are well reproduced
even if $r_{x,s}^2 / \tau_0^2  \approx 0.5$, 
except the $R_{out}$ radius parameter
 at high $m_t$. For pions, the relative error on 
the radius parameters may reach 50 \% for the side and the longitudinal
momentum component, while it may reach 100\% for the out radius component.
Thus the condition of eq. (\ref{e:xsc})  is  substantial,
it is mandatory to check it before concluding about the validity of the 
analytical approximations e.g. in a given data analysis.

\vspace*{1.0 truecm}
\leftline{\bf 14. Departure from Mid-Rapidity} 
\vspace*{0.4 truecm}
Finally we turn our attention to LCMS radius parameters and
IMD results off the mid-rapidity region.
We use such parameters that  the analytical
approximations are precise to $ 10 - 20 $ \% relative errors
at mid-rapidity.  For $ y \ne y_0$, the out-longitudinal cross-term picks up
non-vanishing values. This cross-term must vanish at $m_t = m$,
it develops a large modulus around $m_t - m = 50$ MeV,
which decreases fast with increasing value of $m_t$,
Figures \ref{f:14} - \ref{f:16}.
Surprisingly, the precision of the analytical
 approximations for side, longitudinal
radius parameters and for the IMD is {\it improved} by moving one or
two units away from the mid-rapidity.
This is due to the fact that $r_{x,s}$, $R_*$,
$\Delta\eta_*$ and $\Delta\tau_*$ all decrease with increasing
$\eta_s$ which corresponds to increasing values of 
$\mid y - y_0 \mid$. 
Figure \ref{f:15} indicates that the analytic calculations may remain
reliable even if $y = 1$ for a source centered around $y_0 = 3$
and having a width $\Delta\eta = 1.5$, except the analytical result
for the cross term in the low $m_t$ region.
However, the analytical results
for the cross-term are below the 20 \% relative error limit if
$m_t - m \ge 150 $ MeV or if $\Delta\eta \le 0.7$. With other words,
the deviation 
of the analytical result for the cross-term from its numerically calculated
 value decreases very fast with increasing $m_t$. 
Figures \ref{f:14} - \ref{f:16} indicate, that the analytical results for 
$R_{side}$, $R_{long}$ and the IMD can provide an approximation
with as small as 5 - 10 \% relative error, for $R_{out}$ with 15 \%
relative error even if the departure from mid-rapidity
is substantial, $\mid y - y_0 \mid \simeq 2 \Delta\eta$.
According to Figure \ref{f:16} the analytical approximations are
reliable for $m_t - m > 150 $ MeV if $y = 6$, for lower values
of $m_t$ the longitudinal saddle-point $\eta_s^{LCMS}$ is badly determined,
due to the break-down of the linearized saddle-point  equations.

\vspace*{1.0 truecm}
\leftline{\bf 15. Summary and Conclusions} 
\vspace*{0.4 truecm}

We have presented a detailed numerical testing of the analytical approximations,
as were given in \cite{3d} describing a three-dimensionally expanding,
cylindrically symmetric, finite system. 
We find that the typical precision of the analytical approximations is
characterized by a relative error of about 10 - 20 \%, but
in some cases much more precise agreement is found.
Numerically, we have found that the analytic approximations are
reliable (within 10 -20 \% relative errors) if the transverse
position of the saddle-point satisfies 
$r_{x,s} / \tau_0 \le 0.6$ in some kinematic region.
Numerically we find that this condition is the most sensitive 
among all the analytically found conditions for the validity
of the analytical approximations, and a small violation of this
condition may result in large deviations between the analytical and
numerical values.
This corresponds to the break-down of the linearized saddle-point 
equations in this case. 

Surprisingly, we find that moderate  deviation from mid-rapidity {\it improves}
the agreement between the analytical results and the numerically found
values for the HBT radius parameters and the invariant momentum distribution.
This is due to the decrease of the transverse position of the saddle-point
and the decrease of the effective source sizes with increasing deviation
from mid-rapidity $\mid y - y_0\mid $. For large deviations from mid-rapidity,
e.g. $\mid y - y_0 \mid = 2 \Delta \eta$, the analytical values for the 
cross-term  and for the HBT radius parameters are
found to be unreliable in the $m_t - m \le 150$ MeV interval.
However, for larger values of $m_t - m$ the HBT radius parameters,
including even the cross-term, are reproduced
correctly even at $y = 6$. This lower limit $m_t > 150$ MeV corresponds to the 
requirement $\eta_s^{LCMS} \le 0.9$.

In conclusion, we find that the analytical results are more reliable than
expected before, since they are reliable in a fairly large rapidity
interval. When comparing the analytic results to data,
one has to check if the transverse position  and the space-time rapidity
of the saddle-point is not too large in LCMS.
The essential numerical conditions of consistency are found to be
 $r_{x,s}/\tau_0 < 0.6$ and $\eta_s^{LCMS} \le 0.9$. 

\vspace*{1.0 truecm}
\leftline{\bf Acknowledgments}
\vspace*{0.4 truecm}

The authors would like to thank B. Tom\'a\v{s}ik for his
contributions to the model-testing and
for his careful reading of the manuscript.
Cs. T. would like to thank J\'an Pi\v{s}\'ut and the Organizing Committee 
for invitation and local support at the Bratislava
School and Workshop on Heavy Ion Physics. He thanks
 Gy\"orgyi and  M. Gyulassy for special hospitality at Columbia University.
This work has been supported by the HNSF grants
OTKA - F4019 and W01015107.

\vfill\eject


\leftline{\bf References}
\begin{thebibliography}{99}
\bibitem{na35} 		Th. Alber et al, NA35 Collaboration,
			Phys. Rev. Lett. {\bf 74} (1995) 1303
\bibitem{na44}          H. Beker et al, NA44 Collaboration,
                        Phys. Rev. Lett. {\bf 74} (1995) 3340
\bibitem{na49}          T. Alber et al, NA49 Collaboration,
                        Nucl. Phys. {\bf A590} (1995) 453c
\bibitem{wa93}		S. Slegt et al, WA93 Collaboration,
			Nucl. Phys. {\bf A590} (1995) 469c
\bibitem{nr}
			T. Cs\"org\H o, B. L\"orstad and J. Zim\'anyi,
                        Phys. Lett. {\bf B338} (1994) 134-140
\bibitem{1d} 
			T. Cs\"org\H o, Phys. Lett. {\bf B347} (1995) 354-360
\bibitem{halo}
			T. Cs\"org\H o, B. L\"orstad and J. Zim\'anyi,
			hep-ph/9411307, Z. Phys. C in press
\bibitem{qm95}		T. Cs\"org\H o and B. L\"orstad,
                        hep-ph/9503494,
                        Nucl. Phys. {\bf A590} (1995) 465c
\bibitem{3d} 		T. Cs\"org\H o and B. L\"orstad, 
			CU-TP-717, LUNFD6/(NFFL-7082)-Rev (1994),
			hep-ph/9509213
\bibitem{mpd95}		T. Cs\"org\H o and B. L\"orstad,
			Proc. of XXV-th
			ISMPD, Stara Lesna, 1995 (World Scientific in press,
			ed. L. Sandor et al.),
			 hep-ph/9511404 
\bibitem{sinyu93}
			Yu. Sinyukov, Nucl. Phys. {\bf A566} (1994) 589c-592c;
\bibitem{hhm:te}
			Yu. Sinyukov, in Proc. HHM:TE (Divonne,
                        June 1994, Plenum Press, J. Rafelski et al, eds.)
\bibitem{akkelin}       S. V. Akkelin and Yu. M. Sinyukov,
                        preprint ITP-63-94E
\bibitem{marburg}	 B.R. Schlei, U. Ornik, M. Pl\"umer, D. Strottman
			and R.M. Weiner,
			preprint  LA-UR-95-3232,  hep-ph/9509426
\bibitem{m93}		J. Bolz, U. Ornik, M. Pl\"umer, B. R. Schlei
                        and R. M. Weiner, Phys. Lett. {\bf B300} (1993) 404-409
\bibitem{uli_sh} 	S. Chapman, P. Scotto and U. Heinz, 
			Phys. Rev. Lett. {\bf 74} (1995) 4400
\bibitem{uli_l}           S. Chapman, P. Scotto and U. Heinz,
                        hep-ph/9409349, Heavy Ion Physics {\bf  1} (1995) 1;
\bibitem{nix}		S. Chapman, J. R. Nix and U. Heinz,
                        nucl-th/9505032
\bibitem{wiedemann}     U. A. Wiedemann, P. Scotto and U. Heinz,
                        nucl-th/9508040, Phys. Rev. {\bf C53} (1996) 918
\bibitem{losalamos} 	B. V. Jacak, NA44 Collaboration,
			Nucl. Phys. {\bf A590} 215c;
			D.E. Fields et al. Phys. Rev. {\bf C52}, 986(1995)
\bibitem{humanic}	T. J. Humanic,
			Phys. Rev. {\bf C50} (1994) 2525-2531
\bibitem{sinyukov} 	A. Makhlin and Y. Sinyukov,
                        Z. Phys. {\bf C39}, (1988) 69
\bibitem{pratt_csorgo}  S. Pratt, T. Cs\"org\H o, J. Zim\'anyi, Phys. Rev.
                        {\bf C42} (1990) 2646
\bibitem{zajc} 		W. A. Zajc, in NATO ASI Series {\bf B303}, p. 435
\bibitem{chapman_heinz} S. Chapman and U. Heinz,
                        Phys. Lett. {\bf B340} (1994) 250
\bibitem{deutschmann} 	M. Deutschmann et al,
			Nucl. Phys. {\bf B204} (1982) 333
\bibitem{bertsch} 	G. F. Bertsch, Nucl. Phys. {\bf A498} (1989) 173c
\bibitem{bengt} 	B. L\"orstad,  Int. J. Mod. Phys.
                        {\bf A12} (1989) 2861-2896
\bibitem{ua1} 		N. Neumeister et al, UA1 Collaboration, 
			Act. Phys. Slov.  {\bf 44} (1994) 113-125
\bibitem{LCMS}  	T. Cs\"org\H o and S. Pratt,
                        {\bf KFKI-1991-28/A}, p. 75
\bibitem{cooper}        F. Cooper and G. Frye,
                        Phys. Rev. {\bf D10} (1974) 186
\bibitem{japan} 	B. L\"orstad, hep-ph/9509214
\end{thebibliography}
\end{document}